%% file: paper.tex
\documentclass[12pt,a4paper,dvips]{article}
\usepackage{a4p}
\usepackage{cite,mcite,epsfig}
\usepackage{graphicx}
\usepackage{physics}
\usepackage{l3_title}
\usepackage{ifthen}
\journalname{Phys. Lett. B}
\date{July 25, 2002}
\preprint{2002-062}
\def\ifmath#1{\relax\ifmmode #1\else $#1$\fi}%
\def\rd{\ifmath{{\mathrm{d}}}}
\def\rb{\ifmath{{\mathrm{b}}}}
\def\re{\ifmath{{\mathrm{e}}}}

\def\rZ{\ifmath{{\mathrm{Z}}}}
\def\rW{\ifmath{{\mathrm{W}}}}
\def\rq{\ifmath{{\mathrm{q}}}}

\def\pairs{\ifmath{{\mathrm{pairs}}}}
\def\ev{\ifmath{{\mathrm{ev}}}}

\def\mix{\ifmath{{\mathrm{mix}}}}
\def\MC{\ifmath{{\mathrm{MC}}}}

\def\inter-W{\ifmath{{\mathrm{inter-W}}}}

\def\exp{\ifmath{{\mathrm{exp}}}}

\def\data{\ifmath{{\mathrm{data}}}}
\def\bg{\ifmath{{\mathrm{bg}}}}

\newcommand{\light}{\ensuremath{\mathrm{udcs}}}

\newcommand{\PARJ}[1]{{\scshape parj(#1)}}
\newcommand{\noi}{\ensuremath{\mathrm{no\,inter}}}

\newlength{\capindent}
\newlength{\capwidth}
\newlength{\figwidth}
\newcommand{\icaption}[2][!*!,!]{\hspace*{\capindent}%
  \begin{minipage}{\capwidth}
    \ifthenelse{\equal{#1}{!*!,!}}%
      {\caption{#2}}%
      {\caption[#1]{#2}}
  \end{minipage}}

\newcommand{\pho}{\phantom{0}}

\begin{document}
\begin{titlepage}
\title{Measurement of Bose-Einstein Correlations\\ in {\boldmath $\re^{+}\re^{-}\rightarrow\rW^{+}\rW^{-}$}
Events at LEP}
\author{L3 Collaboration}
%
%
\begin{abstract}
  Bose-Einstein correlations in W-pair production at LEP are investigated
  in a data sample of 629 pb$^{-1}$ collected by the
  L3 detector at $\sqrt{s}=$ 189--209\,\GeV. Bose-Einstein correlations between
  pions within a W decay are observed and found to be in good
  agreement with those in light-quark Z decay. No evidence is found for 
  Bose-Einstein correlations between hadrons coming from different W's in
  the same event.
\end{abstract}
%
%
\submitted
\end{titlepage}
%
%

\section*{Introduction}
In hadronic Z decay, Bose-Einstein correlations (BEC) are observed as an
enhanced production of identical bosons at small
four-momentum difference~\cite{becZ,elong}. 
BEC are also expected within hadronic W decay (intra-W BEC). 
At LEP energies, in fully-hadronic $\rW^{+}\rW^{-}$ events ($\rq\bar{\rq}\rq\bar{\rq}$)
the W decay products overlap in space-time. Therefore, it is also natural
to expect~\cite{Lonn95,BEalg} BEC between identical bosons originating from different W's (inter-W BEC).
A comparison of BEC in fully-hadronic $\rW^{+}\rW^{-}$ events with
those in semi-hadronic $\rW^{+}\rW^{-}$ events ($\rq\bar{\rq}\ell\nu$), serves as a probe to study
inter-W BEC. 
Together with colour reconnection~\cite{colrec,col2},
inter-W BEC form a potential bias in the determination of the W mass at LEP.

In this Letter we search for evidence of inter-W BEC.
We examine all BEC, intra-W as well as inter-W,
and make a comparison with the BEC observed in hadronic Z decay with and without the contribution of
$\rZ\rightarrow\rb\bar{\rb}$ decay.
This Letter updates the analysis of Reference~\citen{l3ww} using about 3.5 times more data.

\section*{Data and Monte Carlo}
The data used in this analysis were collected by the L3 detector~\cite{l3} 
at $\sqrt{s}=189-209\,\GeV$ and correspond to a total integrated luminosity of 629 pb$^{-1}$. 
These data are grouped in seven energy bins listed in Table~\ref{table0}.

Fully-hadronic and semi-hadronic $\rW^{+}\rW^{-}$ events are selected with
criteria similar to those described in Reference~\citen{selww}.
An additional requirement for the fully-hadronic channel is a cut on the neural network output~\cite{selww}
to further separate the signal from the dominant $\re^{+}\re^{-}\rightarrow \rq\bar{\rq}(\gamma)$ background.
In total, about 3,800 semi-hadronic and 5,100 fully-hadronic events are selected.

The event generator KORALW~\cite{koralw} with the BEC algorithm BE$_{32}$~\cite{BEalg}
is used to simulate the signal process. The values of the BE$_{32}$ parameters are found by tuning the 
Monte Carlo (MC) to Z-decay data depleted in b-quark events. Both the BEC and the
fragmentation parameters are tuned simultaneously~\footnote{
The BEC parameter values \PARJ{92}=0.90, giving the BEC strength, and \PARJ{93}=0.43\,\GeV,
the inverse effective source size, are obtained.}. This set is used in this analysis.
Systematic studies are made using an alternative set of parameter values~\footnote{
The alternative set uses the BEC parameter values \PARJ{92}=1.68 and \PARJ{93}=0.38\,\GeV.},
obtained by tuning to Z-decay data of all flavours and used in Reference~\citen{l3ww}.

The background processes $\re^{+}\re^{-}\rightarrow\rq\bar{\rq}(\gamma)$, $\re^{+}\re^{-}\rightarrow\rZ\rZ$
and $\re^{+}\re^{-}\rightarrow\rZ\re^{+}\re^{-}$ are generated using PYTHIA~\cite{pythia}.
For the $\rq\bar{\rq}(\gamma)$ channel KK2f~\cite{kk2f} is also used.  BEC are included in both programs.
The generated events are passed through the L3 detector simulation program~\cite{l3sim}, 
reconstructed and subjected to the same selection criteria as the data.

The selection efficiencies of the 
$\rW^{+}\rW^{-}$ 
channels
$\rq\bar{\rq}\re\nu$, $\rq\bar{\rq}\mu\nu$, $\rq\bar{\rq}\tau\nu$ and $\rq\bar{\rq}\rq\bar{\rq}$
are found to be 83\%, 75\%, 50\%  and 86\%, respectively. 
The purities of these channels
are around 95\%, 95\%, 85\%  and 80\%, respectively, varying a few percent between the different energy bins. 
The selection efficiencies of fully-hadronic
events changes by less than 0.5\% when BEC (intra-W, or both intra-W and inter-W) are not simulated.

A sample of hadronic Z-decay events is selected from the data collected by the L3 detector at
$\sqrt{s}=91.2\;\GeV$,
corresponding to a total integrated luminosity of 48.1 pb$^{-1}$. The event selection, which is similar to that
presented in Reference~\citen{elong}, results in about one million hadronic Z-decay events with a purity greater than
99\%.
Since b quarks are greatly suppressed in W decays, and to compare with the $\rW^{+}\rW^{-}$ data,
a light-quark (u, d, s or c) enhanced Z-decay sample is selected
using an anti-b tagging procedure~\cite{btag} to reduce the $\rb\bar{\rb}$ fraction from 22\% to 3\%. 
This results in about 180,000 Z-decay events depleted in b quarks.

The charged pions used for the BEC study are detected as tracks in the central tracker, using
selection criteria similar to those of Reference~\citen{l3ww}.
About 82\% of the tracks selected in MC samples are pions.
This selection yields about one million pairs of like-sign particles in the fully-hadronic channel 
and about 200,000 pairs in the semi-hadronic channel.

\section*{Analysis Method}
\subsection*{The Two-Particle Bose-Einstein Correlation Function}
BEC are due to interference between identical bosons which are
close to one another in phase space. 
These correlations can be studied in terms of the two-particle correlation function
\begin{equation}
   R_2(p_1,p_2)=\frac{\rho_{2}(p_1,p_2)}{\rho_{0}(p_1,p_2)} \ \ \ ,
\label{r2corr}
\end{equation}
where $\rho_{2}(p_1,p_2)$ is the two-particle number density of particles with four-momenta $p_1$ and $p_2$, and 
$\rho_{0}(p_1,p_2)$ the same density in the absence of BEC. 
As is customary in correlation studies, the normalization is 
$\int\int\rho_2(p_1,p_2)\,\mathrm{d}p_1\mathrm{d}p_2=\langle n(n-\delta_{12})\rangle$,
where $n$ is the number of particles and $\delta_{12}$ is 1 for identical particles and 0 otherwise.
The largest BEC occur at small absolute
four-momentum difference $Q\equiv\sqrt{-(p_1-p_2)^2}$ and $R_2$ is parametrized in this 
one-dimensional distance measure by defining
\begin{equation}
  \rho_{2}(Q)=\frac{1}{N_{\ev}}\frac{\rd n_{\pairs}}{\rd Q} \ \ \ ,
\label{eq0}
\end{equation}
where $N_{\ev}$ is the number of selected events and $n_{\pairs}$ the number of like-sign track pairs in
the $N_{\ev}$ events, retaining the normalization of $\rho_2(p_1,p_2)$.

Apart from the quark flavour, hadronic W and Z decays are expected to be similar. Hence, we measure
the correlation function for a light-quark Z-decay sample, $R_2^{\rZ\rightarrow\light}$,
in order to compare it with those of the fully-hadronic 
$\rW^{+}\rW^{-}$ events, $R_2^{\rW\rW}$, and the semi-hadronic $\rW^{+}\rW^{-}$ events, $R_2^{\rW}$. 

If there is no inter-W interference, we can write~\cite{cwk}
\begin{equation}
  \rho_2^{\rW\rW} (p_1,p_2)=2\rho_2^{\rW}(p_1,p_2)+2\rho_1^{\rW}(p_1)\rho_1^{\rW}(p_2) \ \ \ ,
\label{eqindep2}
\end{equation}
where the assumption is made that the densities for the $\rW^{+}$ and $\rW^{-}$ 
bosons are the same.
In order to transform from $p_1,p_2$ to $Q$,
we replace
the product of the single-particle densities
by a two-particle density, $\rho_{\mix}$, yielding
$ \rho_{2}^{\rW\rW}(Q)= 2\rho_2^{\rW}(Q) + 2\rho_{\mix}^{\rW\rW}(Q)$.
We obtain $\rho_{\mix}^{\rW\rW}(Q)$
by pairing particles
originating from two different semi-hadronic events. By construction, particles in these pairs
are uncorrelated. The event mixing procedure is explained in detail in References~\citen{l3ww} and~\citen{thesis}.
This procedure also takes into account the momentum correlation between the two W's.
The other two densities, 
$\rho_2^{\rW\rW}$ and $\rho_2^{\rW}$, are measured directly
in the fully-hadronic and the semi-hadronic events, respectively.

For comparison to the actual $\rW^{+}\rW^{-}$ data, $R_2^{\rW\rW}$ is estimated for the case that inter-W BEC are absent.
In this case
\begin{equation}
  R_{2,\;\noi}^{\rW\rW}(Q) = \frac{\rho_{2}^{\rW\rW}(Q)}{\rho_{0}^{\rW\rW}(Q)}
                           = \frac{2\rho_2^{\rW}(Q)+2\rho_{\mix}^{\rW\rW}(Q)}{2\rho_{0}^{\rW}(Q)+2\rho_{\mix}^{\rW\rW}(Q)}
                           = 1+\left(1-F(Q)\vphantom{R_2^{\rW}}\right)\left(R_2^{\rW}(Q)-1\right) \ \ \ ,
\label{r2wwvsrw2}
\end{equation}
where
\begin{equation}
   F(Q)\equiv\frac{2\rho_{\mix}^{\rW\rW}(Q)}{\rho_{0}^{\rW\rW}(Q)}
\label{overlapf}
\end{equation}
is the fraction of pairs,
in which the two particles come from different W's,
that would occur 
in fully-hadronic events 
if there were no BEC.
In Equation~(\ref{r2wwvsrw2}) we assume
that $\rho_{\mix}^{\rW\rW}$ is the same for a sample with and without intra-W BEC (and no inter-W BEC).
From a MC study, this assumption is found to be valid.
In the extreme case of complete overlap of the two-particle densities,
$\rho_{0}^{\rW}=\rho_{\mix}^{\rW\rW}=\frac{1}{4}\rho_{0}^{\rW\rW}$, and then $F(Q)=0.5$.
In the other extreme, namely no overlap, $\rho_{\mix}^{\rW\rW}=0$ at $Q=0$, 
so that $F(Q)=0$ there, with the necessary compensation at large $Q$.
Then BEC are equally strong in fully- and semi-hadronic events.
Thus, depending on the degree of overlap, the strength of the BEC
in fully-hadronic events is reduced by up to 50\%
compared to that of semi-hadronic events.
We estimate $F(Q)$ using a MC without BEC.

\subsection*{Direct Search for Inter-W BEC}
The hypothesis that the two W's decay independently can be directly tested using 
Equation~(\ref{eqindep2}). In particular, the following test statistics are
defined as the difference and the ratio of the left- and right-hand side of Equation~(\ref{eqindep2}) in terms of $Q$
\begin{equation}
  \Delta\rho (Q)=\rho_{2}^{\rW\rW}(Q)-2\rho^{\rW}_2(Q)-2\rho_{\mix}^{\rW\rW}(Q)\label{test1}
\end{equation}
and
\begin{equation}
  D(Q)=\frac{\rho_{2}^{\rW\rW}(Q)}{2\rho^{\rW}_2(Q)+2\rho_{\mix}^{\rW\rW}(Q)} \ \ \ .
\label{test2}
\end{equation}
This method gives access to inter-W correlations directly from the data, with no need of  MC \cite{cwk}.

In the absence of inter-W correlations, $\Delta\rho=0$ and $D=1$. To study inter-W BEC, 
deviations from these values are examined at small values of $Q$ for like-sign particles. The influence of other
correlations or potential bias on these quantities is studied by analysing unlike-sign pairs and MC events.

The event mixing procedure could introduce artificial distortions or
not fully account for some correlations other than BEC or some detector effects, causing
a deviation of $\Delta\rho$ from zero or $D$ from unity for data as well as for a
MC without inter-W BEC. These possible effects are reduced by using the double ratio
\begin{equation}
  D'(Q)=\frac{D(Q)_{\data}}{D(Q)_{\MC,\;\noi}} \ \ \ ,
\label{test3}
\end{equation}
where $D(Q)_{\MC,\;\noi}$ is derived from a MC sample without inter-W BEC.

\section*{Results}
\subsection*{Measurement of the Correlation Function}
To obtain the correlation function $R_2$, Equation~(\ref{r2corr}), for the $\rW^{+}\rW^{-}$ events, 
background is subtracted by replacing $\rho_2(Q)$, Equation~(\ref{eq0}), by
\begin{equation}
   \rho_{2}(Q)=\frac{1}{\mathcal{P}\mathnormal{N}_{\ev}}\left( \frac{\rd n}{\rd Q}
                                                              -\frac{\rd n_{\bg}}{\rd Q}\right) \ \ \ ,
\label{eqbg}
\end{equation}
where $\mathcal{P}$ is the purity of the selection and $n_{\bg}$ is the number of pairs of tracks
corresponding to $(1-\mathcal{P})\mathnormal{N}_{\ev}$ background events.
This density is further corrected for detector resolution, acceptance,
efficiency and for particle misidentification with a multiplicative factor derived from MC.
Since no hadrons are identified, this factor is the 
two-pion density found from MC events at generator level divided by the two-particle density
found using all particles after full detector simulation, reconstruction and selection.
For this detector correction, the no inter-W scenario with the BE$_{32}$ algorithm is used.

The reference distribution $\rho_0$ is calculated from a KORALW MC sample without any BEC at generator
level, using only pions. We use a MC distribution rather than unlike-sign particle pairs from
data, as these are strongly affected by the presence of dynamical correlations, such as resonances. 

\begin{figure}
\begin{center}
  \epsfig{figure=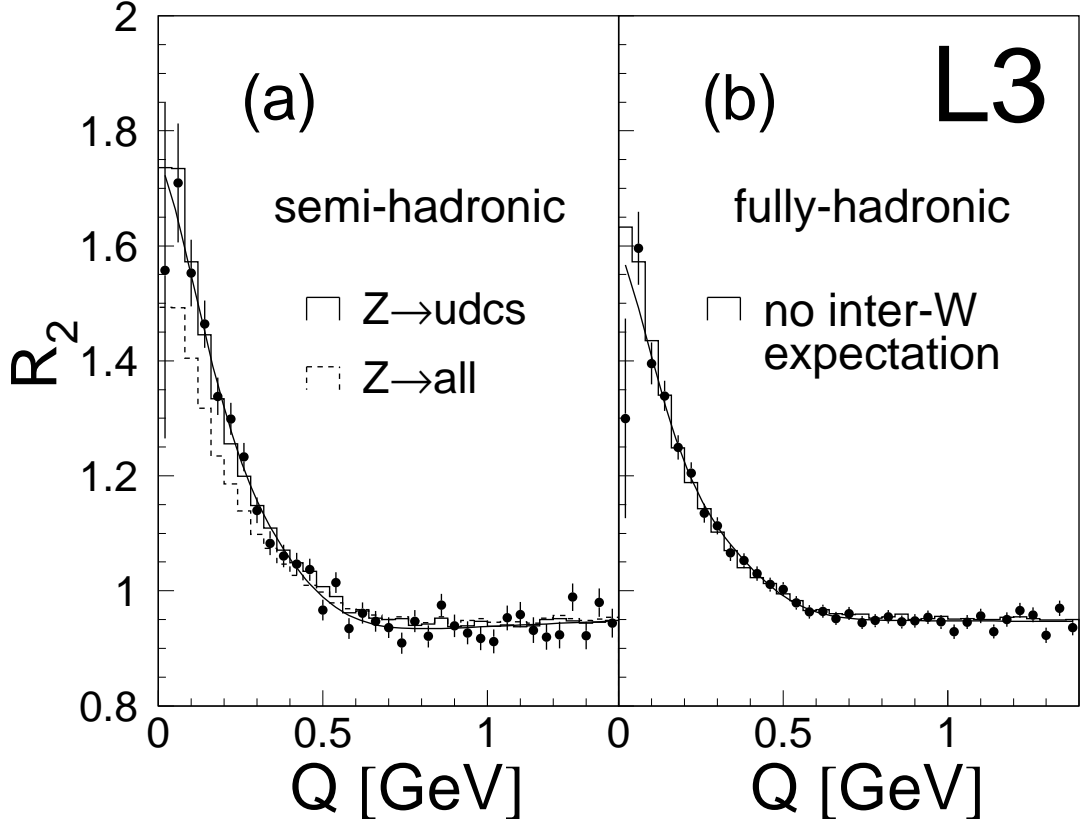, width=\linewidth}
\caption{The correlation function $R_2$ at $\sqrt{s}$ = 189--209\,\GeV, for
(a) semi-hadronic events
and
(b) fully-hadronic events.
The solid curves are the results from the fits of Equation~(\ref{fitwwedge}) to
the $\rW^{+}\rW^{-}$ data. 
In (a) the full histogram is for the light-quark Z-decay
data sample and the dashed histogram is for the data sample containing all hadronic Z decays.
In (b) the full histogram gives the expectation when inter-W BEC are absent,
Equation~(\ref{r2wwvsrw2}), using 
$R_2^{\rW}=R_2^{\rZ\rightarrow\light}$.
}
\label{fig1} 
\end{center}
\end{figure}

Figure~\ref{fig1} shows the correlation function for 
semi-hadronic and fully-hadronic events. 
Results for different energy bins are compatible with each other,
and therefore are combined.

The correlation function of the light-quark Z-decay data sample, $R_2^{\rZ\rightarrow\light}$,
is determined similarly to $R_2^{\rW}$ and $R_2^{\rW\rW}$. Figure~\ref{fig1}a compares
$R_2^{\rZ\rightarrow\light}$ with $R_2^{\rW}$ and, as expected, good agreement is observed.
If b-quark decays of the Z are not removed from the sample,
a depletion of the correlation function is observed at low $Q$, and a clear discrepancy exists
with the W data, as also shown in Figure~\ref{fig1}a.


The BEC enhancement at
low $Q$ values is fitted from $Q=0$ to $1.4\,\GeV$ with the first-order 
Edgeworth expansion of the Gaussian~\cite{edge,elong,l3be3},
\begin{equation}
   R_{2}(Q)=\gamma (1+\delta Q)\left[ 1+\lambda\,\exp\left(-R^2 Q^2\right)
         \left( 1+\frac{\kappa}{6}H_{3}\left(\sqrt{2}RQ\right)\right)\right] \ \ \ ,
\label{fitwwedge}
\end{equation}
using the full covariance matrix~\cite{thesis}, which takes into account the effect
of bin-to-bin correlations.
The parameter $\gamma$ is an overall normalization factor, the term $(1+\delta Q)$ takes into
account possible long-range momentum correlations, $\lambda$ measures the strength of the BEC, $R$ is
related to the effective source size in space-time, $\kappa$ measures the
deviation from the Gaussian and $H_{3}(x)\equiv x^{3}-3x$ 
is the third Hermite polynomial.
The fit results are given in Table~\ref{table1} and shown in Figure~\ref{fig1}. 

BEC are observed ($\lambda>0$) in both fully-hadronic and semi-hadronic events. The value 
of $\lambda$ is higher for the semi-hadronic than for the fully-hadronic channel,
which could be due to
a partial suppression of inter-W BEC  ~\cite{cwk} combined with
incomplete overlap of $\rho_2^{\rW^+}$ and $\rho_2^{\rW^-}$.
Using a MC without BEC to correct the data for detector effects, reduces $\lambda$ in both channels by about 30\%,
but the difference in the $\lambda$ values remains.

Different sources of systematic uncertainties on the fit parameters are studied. Track and event selections are varied,
the background fractions are varied by $\pm$20\%, covering biases in the $Q$ distributions of background events,
and different MC's, using different sets of MC parameter values, are used to generate the background events. 
Moreover, the stability of the results is tested by changing the fully-hadronic event selection such that the background
fraction is reduced by more than a factor two.

The influence of the choice of the MC used for
the reference sample and for the detector correction is also taken into account.
For the detector correction a MC with intra-W BEC is used, since
it agrees better with the $Q$ distribution of the raw data than a MC without BEC \cite{l3ww}.
Both sets of parameter values in BE$_{32}$ are used and for the
fully-hadronic channel both the no inter-W and inter-W scenarios are considered. For the inter-W
scenario, the systematic uncertainty is estimated using the fraction of inter-W BEC consistent
with the measurements of $\Delta\rho$, $D$ and $D'$ discussed below. 
Finally, a systematic uncertainty is estimated by varying the fit range by $\pm 320\,\MeV$.
The contributions to the systematic uncertainty of $\lambda$ are shown in Table~\ref{table2}.

The correlation function for fully-hadronic events expected for the case of no inter-W BEC,
$R^{\rW\rW}_{2,\;\noi}$, is estimated from Equation~(\ref{r2wwvsrw2}). In this calculation,
the value of $F(Q)$, from Equation~(\ref{overlapf}) using KORALW without BEC, is around 0.2
at low $Q$ and increases to 0.6 for $Q>3\,\GeV$. Using these values of $F(Q)$ and
$R_2^{\rW}=R_2^{\rZ\rightarrow\light}$ in Equation~(\ref{r2wwvsrw2}),
we obtain the no inter-W prediction shown in Figure~\ref{fig1}b as the full histogram.
Good agreement is observed between this distribution and the fully-hadronic $\rW^{+}\rW^{-}$ data, 
thus indicating no or only weak inter-W BEC.

\subsection*{Measurement of {\boldmath $\Delta\rho$}, {\boldmath $D$} and {\boldmath $D'$}, Direct Search for Inter-W BEC}
\begin{figure}
\begin{center}
  \epsfig{figure=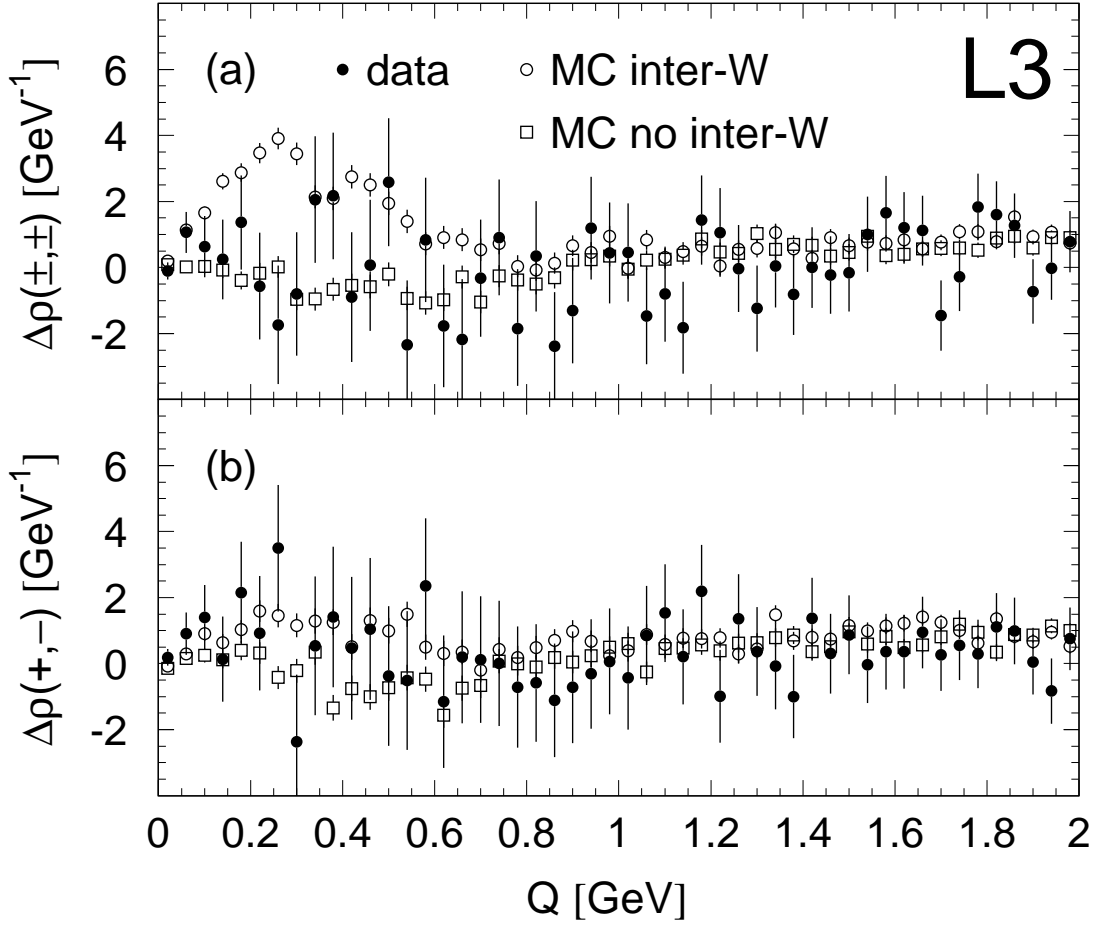, width=1.00\linewidth}
\caption{Distributions for uncorrected data at $\sqrt{s}$ = 189--209\,\GeV\ of (a)
$\Delta\rho(\pm,\pm)$ and (b) $\Delta\rho(+,-)$. Also shown are the MC predictions of KORALW with
and without inter-W BEC.
}
\label{fig2}  
\end{center}
\end{figure}

\begin{figure}
\begin{center}
  \epsfig{figure=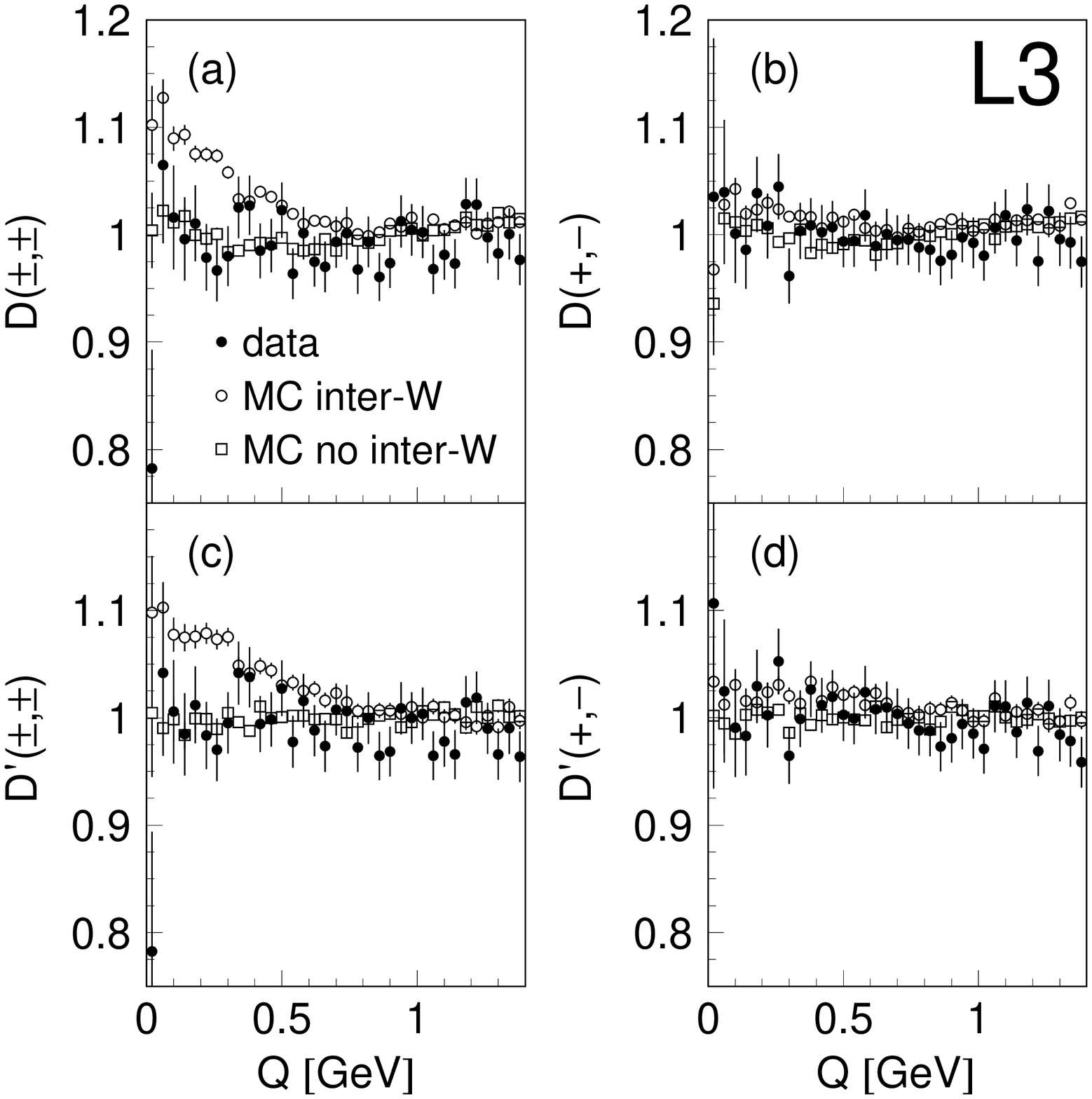, width=1.00\linewidth}
\caption{Distributions for uncorrected data at $\sqrt{s}=$ 189--209\,\GeV\ of
(a) $D(\pm,\pm)$, (b) $D(+,-)$, (c) $D'(\pm,\pm)$ and (d) $D'(+,-)$. 
Also shown are the MC predictions of KORALW with and without inter-W BEC.
}
\label{fig3}  
\end{center}
\end{figure}

Figure~\ref{fig2} shows the distribution of $\Delta\rho$, Equation~(\ref{test1}), for like-sign, $(\pm,\pm)$,
and for unlike-sign, $(+,-)$, particle pairs. Figure~\ref{fig3} 
shows the distributions of $D$ and $D'$, Equations~(\ref{test2}) and~(\ref{test3}), for like-sign and unlike-sign pairs.
For the double ratio $D'$ we use the no inter-W scenario of KORALW as the reference sample.
The distributions of $\Delta\rho$, $D$ and $D'$ are not
corrected for detector effects, but background is estimated from MC and subtracted according to Equation~(\ref{eqbg}),
from $\rho_2^{\rW}$ and $\rho_2^{\rW\rW}$. Also shown in Figures~\ref{fig2}
and~\ref{fig3} are the predictions of KORALW
after full detector simulation, reconstruction and selection. Both the inter-W and no inter-W scenarios are shown.

The inter-W scenario shows an enhancement at small values of $Q$
in the $\Delta\rho$, $D$ and $D'$ distributions for like-sign pairs. 
We also observe a small enhancement for unlike-sign pairs due to the MC implementation of BEC, which shifts
the vector momentum of particles, affecting both the like-sign and unlike-sign particle spectra.
The no inter-W scenario describes the $\Delta\rho(\pm,\pm)$, $D(\pm,\pm)$ and
$D'(\pm,\pm)$ distributions, while the inter-W scenario is disfavoured.

For quantitative comparisons, the integral
\begin{equation}
  J(\pm,\pm)\equiv\int_{0}^{Q_{\max}}\Delta\rho (Q)\,\rd Q
\label{eqnint}
\end{equation}
is computed. Also, the $D'(\pm,\pm)$ distribution is fitted
from $Q=0$ to $1.4\,\GeV$, using the full covariance matrix, with
the parametrization
\begin{equation}
  D'(Q)=(1+\delta Q)(1+\Lambda\exp (-k^2 Q^2)) \ \ \ ,
\label{fitbiglam}
\end{equation}
where $\delta$, $\Lambda$ and $k$ are the fit parameters. 
Both $J(\pm,\pm)$ and $\Lambda$ measure the strength of inter-W BEC. 

The systematic uncertainties on $J(\pm,\pm)$ and on $\Lambda$ are 
listed in Tables~\ref{table3} and~\ref{table4}, respectively. 
In addition to the track and event selections, the amount of background is also varied and
different MC's, using both sets of MC parameter values, are used to generate the background events.
Furthermore, contributions to the systematic uncertainty on $\Lambda$
are obtained by varying the choice of MC for the reference sample in $D'$ using PYTHIA and
KORALW, both with no BEC at all and with only intra-W BEC. MC's without BEC are used to estimate
the effect of residual intra-W BEC. The effect of various models for colour
reconnection~\footnote{The so-called SKI (with reconnection probability of about 30\%), 
SKII and SKII'~\cite{col2} models, as implemented in PYTHIA,
are used.} is included. Changes in the fit range ($\pm 400\,\MeV$), in the bin size (from 40 to 80 $\MeV$)
and in the parametrization of Equation~(\ref{fitbiglam}) (removing $(1+\delta Q)$ from the fit) 
also contribute to the systematic uncertainty on $\Lambda$.

In the mixing procedure, a semi-hadronic event is allowed to be combined with all possible other semi-hadronic 
events. To be sure that this does not introduce a bias, the analysis is repeated for a mixed sample where every
semi-hadronic event is used at most once. The influence of the mixing procedure is also studied by not only combining
oppositely charged W's, but also like-sign W's. The influence of an extra momentum~\cite{l3ww},
used in the event mixing, is also included as a systematic effect. The effect of these
three changes in the mixing procedure is also given in Tables~\ref{table3} and~\ref{table4}. Moreover,
the analysis is repeated removing the cut on the neural network output for the mixed events.
Furthermore, the effect of uncertainties in the energy calibration of the calorimeters is studied. Finally, the
influence of incorrect assignment of tracks to $\tau$ or $\rq\bar{\rq}$ systems
in the $\rq\bar{\rq}\tau\nu$ channel is investigated.

\begin{figure}
\begin{center}
\begin{tabular}{c@{\hspace{-1.5cm}}c}
\hspace{-1.6cm}
  \epsfig{figure=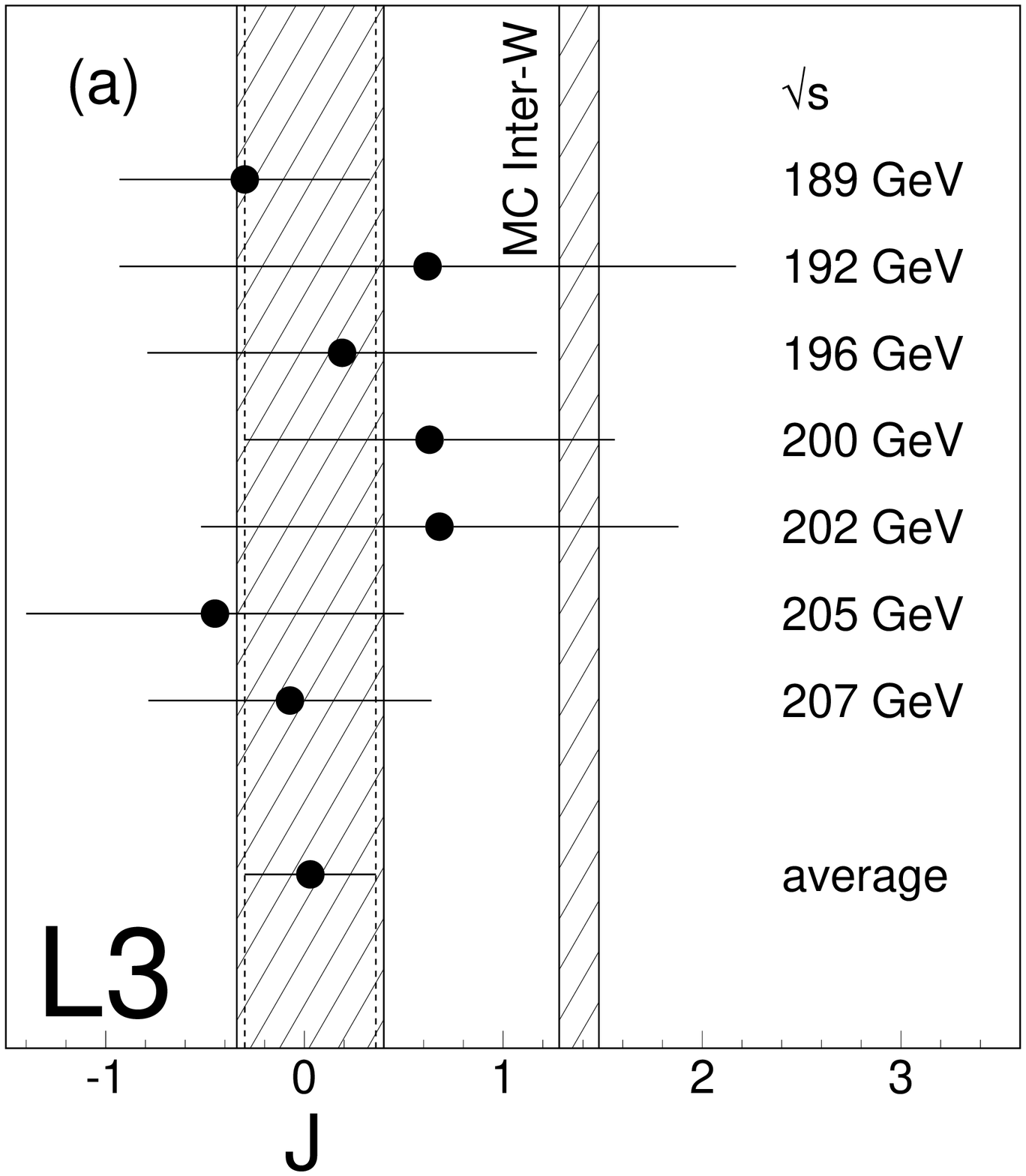, height=10.3cm,width=.6\linewidth} &
  \epsfig{figure=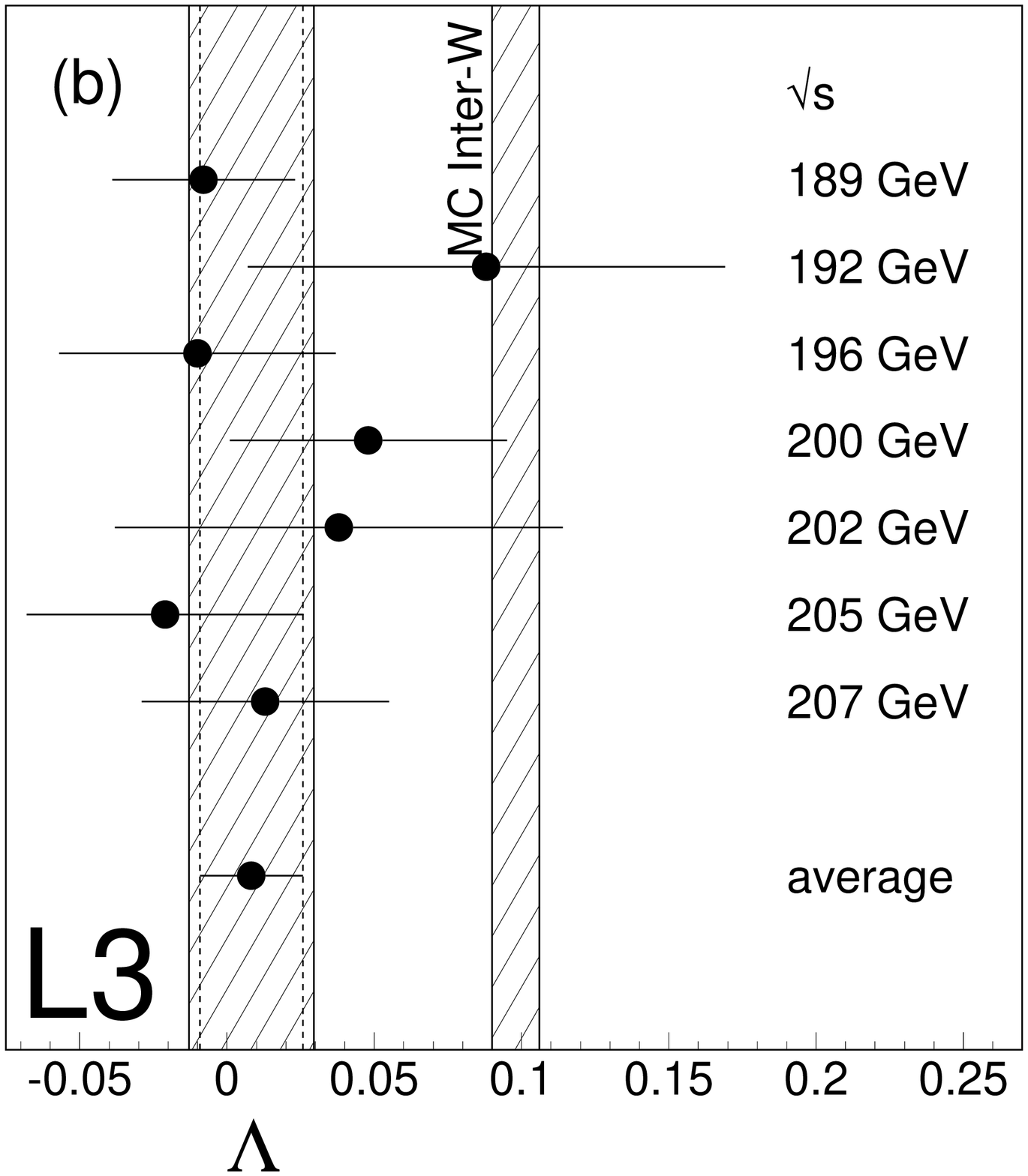, height=10.3cm,width=.6\linewidth}
\end{tabular}
\caption{Values of (a) the integral $J(\pm,\pm)$ and (b) the $\Lambda$ parameter, at different centre-of-mass
energies and their average. The uncertainties are statistical only. The wide bands show the average value of the data including
the systematic uncertainties. Also shown are the MC predictions of KORALW with inter-W BEC. 
}
\label{fig4} 
\end{center}
\end{figure}

The value of $J(\pm,\pm)$ is computed using the full covariance matrix,
taking $Q_{\max}=0.68$\,\GeV, the value where the two MC scenarios have converged to less than one
standard deviation. 
The results for each centre-of-mass energy, displayed in Figure~\ref{fig4}a,
are consistent with each other.
Combining all $J(\pm,\pm)$ values results in 
   $$J(\pm,\pm)=0.03\pm 0.33 \pm 0.15 \ \ \ , $$
where the first uncertainty is statistical and the second systematic.
Using KORALW with the inter-W scenario gives $J(\pm,\pm)=1.38\pm 0.10$, where
the uncertainty is statistical only.
In Figure~\ref{fig4}a this value is shown as a vertical band. It 
disagrees with the value of the data by 3.6 standard deviations.
For unlike-sign pairs we obtain $J(+,-)=0.01\pm 0.36\pm 0.16$, consistent with zero.

The value of the fit parameter $\Lambda$, Equation~(\ref{fitbiglam}),
is shown in Figure~\ref{fig4}b for each energy bin. 
Combining all $\Lambda$ values results in 
  $$\Lambda=0.008\pm 0.018\pm 0.012 \ \ \ , $$
where the first uncertainty is statistical and the second systematic. The value of $k$ is found to
be $0.4\pm0.4\pm0.3$ fm and the correlation coefficient between $\Lambda$ and $k$ is 0.45.
A similar fit is performed for the KORALW MC sample with inter-W BEC, resulting in
$\Lambda=0.098\pm0.008$,
where the uncertainty is statistical only. In Figure~\ref{fig4}b this value is shown as a vertical band. 
It disagrees with the value of the data by 3.8 standard deviations.
Using the alternative set of MC parameters results in $J(\pm,\pm)=1.78\pm0.10$ and $\Lambda=0.126\pm 0.008$,
where the uncertainties are statistical only. 

To summarize, an excess at small values of $Q$ in the distributions of $\Delta\rho (\pm,\pm)$, $D(\pm,\pm)$
and $D'(\pm,\pm)$ is expected from inter-W BEC, but none is seen. These distributions agree well with
KORALW using BE$_{32}$ without inter-W BEC, but not when inter-W BEC are included. We thus find no evidence
for BEC between identical pions originating from different W's.

\section*{Acknowledgements}
Clarifying discussions with \v{S}. Todorova-Nov\'a are gratefully acknowledged.

\newpage
\input namelist256.tex

\begin{table}
\centering

\vspace{0.3cm}
\begin{tabular}{|l|c|c|c|c|c|c|c|}
\hline
$\langle\sqrt{s}\rangle$ (\GeV) & 188.6 & 191.6 & 195.5 & 199.5 & 201.8 & 204.8 & 206.6 \\ \hline
Integrated luminosity (pb$^{-1}$)   & 176.8 & \phantom{0}29.8 & \phantom{0}84.1 & 
\phantom{0}83.3 & \phantom{0}37.2 & \phantom{0}79.0 & 139.1 \\
\hline
\end{tabular}
\caption{Average centre-of-mass energies and corresponding integrated luminosities.}
\label{table0}
\end{table}

\begin{table}
\centering

\vspace{0.3cm}
\begin{tabular}{|c|c|c|}
\hline
Parameter & Fully-hadronic & Semi-hadronic \\
\hline
$\gamma$               & $0.92\pm0.02\pm0.02$ & $0.91\pm0.02\pm0.02$  \\
$\lambda$              & $0.74\pm0.04\pm0.06$ & $0.95\pm0.08\pm0.05$  \\
$R$ (fm)               & $0.73\pm0.03\pm0.06$ & $0.71\pm0.04\pm0.07$  \\
$\delta$ (\GeV$^{-1}$) & $0.00\pm0.02\pm0.05$ & $0.03\pm0.02\pm0.05$  \\
$\kappa$               & $0.80\pm0.09\pm0.10$ & $0.61\pm0.14\pm0.11$  \\
$\chi^2$/dof           & 27.2/30              & 35.4/30               \\
\hline
\end{tabular}
\caption{
  Values of the fit parameters $\gamma$, $\lambda$, $R$, $\delta$ and $\kappa$
  for the fully-hadronic and semi-hadronic events.
  The first uncertainty is statistical, the second systematic.
  The $\chi^2$ and number of degrees of freedom (dof) are also given.
}
\label{table1}
\end{table}

\begin{table}
\centering

\vspace{0.3cm}
\begin{tabular}{|l|c|c|}
\hline
Source & Fully-hadronic & Semi-hadronic \\
\hline
Track selection          & 0.021 & 0.016 \\
Event selection          & 0.013 & 0.014 \\
Background contribution  & 0.018 & 0.021 \\
Alternative MC as reference & 0.013 & 0.013 \\
Alternative MC for correction & 0.022 & 0.024 \\
Inter-W BE in MC correction & 0.032 &   -   \\
Fit range                & 0.020 & 0.018 \\
\hline
Total                    & 0.06\phantom{0} & 0.05\phantom{0} \\
\hline
\end{tabular}
\caption{Contributions to the systematic uncertainty on the $\lambda$ parameter for the
  fully-hadronic and semi-hadronic events.}
\label{table2}
\end{table}

\begin{table}
\centering

\vspace{0.3cm}
\begin{tabular}{|l|c|}
\hline
Source & Contribution \\
\hline
Track selection               & 0.084 \\
Event selection               & 0.068 \\
Background contribution       & 0.055 \\
Mixing procedure              & 0.065 \\
Neural network cut            & 0.038 \\
Energy calibration            & 0.024 \\
Track misassignment in $\rq\bar{\rq}\tau\nu$ channel      & 0.038 \\
\hline
Total                         & 0.15\pho \\
\hline
\end{tabular}
\caption{Contributions to the systematic uncertainty of $J(\pm,\pm)$.}
\label{table3}
\end{table}

\begin{table}
\centering

\vspace{0.3cm}
\begin{tabular}{|l|c|}
\hline
Source & Contribution \\
\hline
Track selection                          & 0.0029 \\
Event selection                          & 0.0049 \\
Background contribution                  & 0.0042 \\
Alternative MC as a reference            & 0.0060 \\
Colour reconnection                      & 0.0026 \\
Fit range                                & 0.0018 \\
Rebinning                                & 0.0020 \\
Fit parametrization                      & 0.0017 \\
Mixing procedure                         & 0.0044 \\
Neural network cut                       & 0.0033 \\
Energy calibration                       & 0.0017 \\
Track misassignment in $\rq\bar{\rq}\tau\nu$ channel  & 0.0022 \\
\hline
Total                                    & 0.012\pho \\
\hline
\end{tabular}
\caption{Contributions to the systematic uncertainty of the $\Lambda$ parameter.}
\label{table4}
\end{table}

\newpage

\end{document}

%% file: namelist256.tex
\typeout{   }     
\typeout{Using author list for paper 256 -  }
\typeout{$Modified: Jul 15 2001 by smele $}
\typeout{!!!!  This should only be used with document option a4p!!!!}
\typeout{   }
%
%
%
%
%
%

\newcount\tutecount  \tutecount=0
\def\tutenum#1{\global\advance\tutecount by 1 \xdef#1{\the\tutecount}}
\def\tute#1{$^{#1}$}
\tutenum\aachen            
\tutenum\nikhef            
\tutenum\mich              
\tutenum\lapp              
\tutenum\basel             
\tutenum\lsu               
\tutenum\beijing           
\tutenum\berlin            
\tutenum\bologna           
\tutenum\tata              
\tutenum\ne                
\tutenum\bucharest         
\tutenum\budapest          
\tutenum\mit               
\tutenum\panjab            
\tutenum\debrecen          
\tutenum\dublin            
\tutenum\florence          
\tutenum\cern              
\tutenum\wl                
\tutenum\geneva            
\tutenum\hefei             
\tutenum\lausanne          
\tutenum\lyon              
\tutenum\madrid            
\tutenum\florida           
\tutenum\milan             
\tutenum\moscow            
\tutenum\naples            
\tutenum\cyprus            
\tutenum\nymegen           
\tutenum\caltech           
\tutenum\perugia           
\tutenum\peters            
\tutenum\cmu               
\tutenum\potenza           
\tutenum\prince            
\tutenum\riverside         
\tutenum\rome              
\tutenum\salerno           
\tutenum\ucsd              
\tutenum\sofia             
\tutenum\korea             
\tutenum\purdue            
\tutenum\psinst            
\tutenum\zeuthen           
\tutenum\eth               
\tutenum\hamburg           
\tutenum\taiwan            
\tutenum\tsinghua          

{
\parskip=0pt
\noindent
{\bf The L3 Collaboration:}
\ifx\selectfont\undefined
 \baselineskip=10.8pt
 \baselineskip\baselinestretch\baselineskip
 \normalbaselineskip\baselineskip
 \ixpt
\else
 \fontsize{9}{10.8pt}\selectfont
\fi
\medskip
\tolerance=10000
\hbadness=5000
\raggedright
\hsize=162truemm\hoffset=0mm
\def\r{\rlap,}
\noindent

P.Achard\r\tute\geneva\ 
O.Adriani\r\tute{\florence}\ 
M.Aguilar-Benitez\r\tute\madrid\ 
J.Alcaraz\r\tute{\madrid,\cern}\ 
G.Alemanni\r\tute\lausanne\
J.Allaby\r\tute\cern\
A.Aloisio\r\tute\naples\ 
M.G.Alviggi\r\tute\naples\
H.Anderhub\r\tute\eth\ 
V.P.Andreev\r\tute{\lsu,\peters}\
% ...
\newpage
\begin{list}{A}{\itemsep=0pt plus 0pt minus 0pt\parsep=0pt plus 0pt minus 0pt
                \topsep=0pt plus 0pt minus 0pt}
\item[\aachen]
 I. Physikalisches Institut, RWTH, D-52056 Aachen, FRG$^{\S}$\\
 III. Physikalisches Institut, RWTH, D-52056 Aachen, FRG$^{\S}$
\item[\nikhef] National Institute for High Energy Physics, NIKHEF, 
     and University of Amsterdam, NL-1009 DB Amsterdam, The Netherlands
\item[\mich] University of Michigan, Ann Arbor, MI 48109, USA
\item[\lapp] Laboratoire d'Annecy-le-Vieux de Physique des Particules, 
     LAPP,IN2P3-CNRS, BP 110, F-74941 Annecy-le-Vieux CEDEX, France
\item[\basel] Institute of Physics, University of Basel, CH-4056 Basel,
     Switzerland
\item[\lsu] Louisiana State University, Baton Rouge, LA 70803, USA
\item[\beijing] Institute of High Energy Physics, IHEP, 
  100039 Beijing, China$^{\triangle}$ 
\item[\berlin] Humboldt University, D-10099 Berlin, FRG$^{\S}$
\item[\bologna] University of Bologna and INFN-Sezione di Bologna, 
     I-40126 Bologna, Italy
\item[\tata] Tata Institute of Fundamental Research, Mumbai (Bombay) 400 005, India
\item[\ne] Northeastern University, Boston, MA 02115, USA
\item[\bucharest] Institute of Atomic Physics and University of Bucharest,
     R-76900 Bucharest, Romania
\item[\budapest] Central Research Institute for Physics of the 
     Hungarian Academy of Sciences, H-1525 Budapest 114, Hungary$^{\ddag}$
\item[\mit] Massachusetts Institute of Technology, Cambridge, MA 02139, USA
\item[\panjab] Panjab University, Chandigarh 160 014, India.
\item[\debrecen] KLTE-ATOMKI, H-4010 Debrecen, Hungary$^\P$
\item[\dublin] Department of Experimental Physics,
  University College Dublin, Belfield, Dublin 4, Ireland
\item[\florence] INFN Sezione di Firenze and University of Florence, 
     I-50125 Florence, Italy
\item[\cern] European Laboratory for Particle Physics, CERN, 
     CH-1211 Geneva 23, Switzerland
\item[\wl] World Laboratory, FBLJA  Project, CH-1211 Geneva 23, Switzerland
\item[\geneva] University of Geneva, CH-1211 Geneva 4, Switzerland
\item[\hefei] Chinese University of Science and Technology, USTC,
      Hefei, Anhui 230 029, China$^{\triangle}$
\item[\lausanne] University of Lausanne, CH-1015 Lausanne, Switzerland
\item[\lyon] Institut de Physique Nucl\'eaire de Lyon, 
     IN2P3-CNRS,Universit\'e Claude Bernard, 
     F-69622 Villeurbanne, France
\item[\madrid] Centro de Investigaciones Energ{\'e}ticas, 
     Medioambientales y Tecnol\'ogicas, CIEMAT, E-28040 Madrid,
     Spain${\flat}$ 
\item[\florida] Florida Institute of Technology, Melbourne, FL 32901, USA
\item[\milan] INFN-Sezione di Milano, I-20133 Milan, Italy
\item[\moscow] Institute of Theoretical and Experimental Physics, ITEP, 
     Moscow, Russia
\item[\naples] INFN-Sezione di Napoli and University of Naples, 
     I-80125 Naples, Italy
\item[\cyprus] Department of Physics, University of Cyprus,
     Nicosia, Cyprus
\item[\nymegen] University of Nijmegen and NIKHEF, 
     NL-6525 ED Nijmegen, The Netherlands
\item[\caltech] California Institute of Technology, Pasadena, CA 91125, USA
\item[\perugia] INFN-Sezione di Perugia and Universit\`a Degli 
     Studi di Perugia, I-06100 Perugia, Italy   
\item[\peters] Nuclear Physics Institute, St. Petersburg, Russia
\item[\cmu] Carnegie Mellon University, Pittsburgh, PA 15213, USA
\item[\potenza] INFN-Sezione di Napoli and University of Potenza, 
     I-85100 Potenza, Italy
\item[\prince] Princeton University, Princeton, NJ 08544, USA
\item[\riverside] University of Californa, Riverside, CA 92521, USA
\item[\rome] INFN-Sezione di Roma and University of Rome, ``La Sapienza",
     I-00185 Rome, Italy
\item[\salerno] University and INFN, Salerno, I-84100 Salerno, Italy
\item[\ucsd] University of California, San Diego, CA 92093, USA
\item[\sofia] Bulgarian Academy of Sciences, Central Lab.~of 
     Mechatronics and Instrumentation, BU-1113 Sofia, Bulgaria
\item[\korea]  The Center for High Energy Physics, 
     Kyungpook National University, 702-701 Taegu, Republic of Korea
\item[\purdue] Purdue University, West Lafayette, IN 47907, USA
\item[\psinst] Paul Scherrer Institut, PSI, CH-5232 Villigen, Switzerland
\item[\zeuthen] DESY, D-15738 Zeuthen, 
     FRG
\item[\eth] Eidgen\"ossische Technische Hochschule, ETH Z\"urich,
     CH-8093 Z\"urich, Switzerland
\item[\hamburg] University of Hamburg, D-22761 Hamburg, FRG
\item[\taiwan] National Central University, Chung-Li, Taiwan, China
\item[\tsinghua] Department of Physics, National Tsing Hua University,
      Taiwan, China
\item[\S]  Supported by the German Bundesministerium 
        f\"ur Bildung, Wissenschaft, Forschung und Technologie
\item[\ddag] Supported by the Hungarian OTKA fund under contract
numbers T019181, F023259 and T037350.
\item[\P] Also supported by the Hungarian OTKA fund under contract
  number T026178.
\item[$\flat$] Supported also by the Comisi\'on Interministerial de Ciencia y 
        Tecnolog{\'\i}a.
\item[$\sharp$] Also supported by CONICET and Universidad Nacional de La Plata,
        CC 67, 1900 La Plata, Argentina.
\item[$\triangle$] Supported by the National Natural Science
  Foundation of China.
\end{list}
}
\vfill
